# SoftSeg: Advantages of soft versus binary training for image segmentation

**Authors**
Charley Gros [1, *], Andreanne Lemay [1,*], Julien Cohen-Adad [1, 2]

**Affiliations**
*1 - NeuroPoly Lab, Institute of Biomedical Engineering, Polytechnique Montreal, Montreal, QC, Canada*
*2 - Functional Neuroimaging Unit, CRIUGM, Université de Montréal, Montreal, QC, Canada*

\* *These authors equally contributed to this work.*

**CORRESPONDING AUTHOR**
Julien Cohen-Adad
Dept. Genie Electrique, L5610
Ecole Polytechnique
2900 Edouard-Montpetit Bld
Montreal, QC, H3T 1J4, Canada
Phone: 514 340 5121 (office: 2264);  Skype: jcohenadad; e-mail: jcohen@polymtl.ca

**ABBREVIATIONS:**
BraTS: brain tumor segmentation
GT: ground truth
MS: multiple sclerosis
MSE: mean squared error
PVE: partial volume effect
ReLU: rectified linear function
RPI: right-to-left, posterior-to-anterior, inferior-to-superior
SCGM: spinal cord gray matter

**Abstract**

Most image segmentation algorithms are trained on binary masks formulated as a classification task per pixel. However, in applications such as medical imaging, this "black-and-white" approach is too constraining because the contrast between two tissues is often ill-defined, i.e., the voxels located on objects' edges contain a mixture of tissues (a partial volume effect). Consequently, assigning a single "hard" label can result in a detrimental approximation. Instead, a soft prediction containing non-binary values would overcome that limitation. In this study, we introduce SoftSeg, a deep learning training approach that takes advantage of soft ground truth labels, and is not bound to binary predictions. SoftSeg aims at solving a regression instead of a classification problem. This is achieved by using (i) no binarization after preprocessing and data augmentation, (ii) a normalized ReLU final activation layer (instead of sigmoid), and (iii) a regression loss function (instead of the traditional Dice loss). We assess the impact of these three features on three open-source MRI segmentation datasets from the spinal cord gray matter, the multiple sclerosis brain lesion, and the multimodal brain tumor segmentation challenges. Across multiple cross-validation iterations, SoftSeg outperformed the conventional approach, leading to an increase in Dice score of 2.0% on the gray matter dataset (p=0.001), 3.3% for the brain lesions, and 6.5% for the brain tumors. SoftSeg produces consistent soft predictions at a tissues' interfaces and shows an increased sensitivity for small objects (e.g., multiple sclerosis lesions). The richness of soft labels could represent the inter-expert variability, the partial volume effect, and complement the model uncertainty estimation, which is typically unclear with binary predictions. The developed training pipeline can easily be incorporated into most of the existing deep learning architectures. It is already implemented in the freely-available deep learning toolbox ivadomed (https://ivadomed.org).

# 1. Introduction

Medical image analysis is at a turning point as a growing number of clinical studies are fully embracing automated processing, thanks to the recent ground-breaking performances of deep learning (De Fauw et al., 2018; Esteva et al., 2017; Litjens et al., 2017). A popular medical application of deep learning is image segmentation, whereby voxels are assigned a label (e.g., 1 if pertaining to the tissue of interest, 0 otherwise). This binary approach to tissue classification is limited in that it does not allow the model to exploit the rich information present in the expert annotation or in the input image. This richness could take the form of inter-expert representation (in case a ground truth is created by several experts) (Carass et al., 2020), level of uncertainty (e.g., a ground truth could take the value 0.5 instead of 1, if the expert is unsure a voxel belongs to a lesion) (T. Nair et al., 2020), or partial volume effect (PVE) (Chaves et al., 2020). PVE is characterized by the mixing of signals coming from different tissue types, and usually happens at their interfaces. For example, if tissue A has the intensity 50 on a MRI scan and tissue B the intensity 100, voxels at their interface exhibit values between 50 and 100, depending on the volume fraction occupied by each tissue. PVE is a well-known problem in computer vision, and it can notably be handled by Gaussian mixture modeling to estimate the true fraction of underlying tissue signals (Lévy et al., 2015; Tohka et al., 2004). However, when it comes to tissue segmentation, PVE is rarely accounted for in conventional deep learning segmentation methods (Akkus et al., 2017; Baumgartner et al., 2019; Billot et al., 2020). Instead, most segmentation pipelines are trained on binary data, with value 0 (outside the tissue) or 1 (inside the tissue), and therefore produce uncalibrated output probabilities. Ideally, segmentation methods would encode predictions as "50 shades of gray", representing partial volume information of the segmented tissue. With this challenge, there is a strong rationale for inputting/outputting "soft" labels in a deep learning segmentation pipeline to better calibrate the model confidence.

## 1.1. Related works

Soft labels have led to a better generalization, faster learning speed, and mitigation of network over-confidence (Müller et al., 2019). Label smoothing was investigated in image classification (Pham et al., 2019; Szegedy et al., 2016), style transfer (Zhao et al., 2020), speech recognition (Chorowski & Jaitly, 2016), and language translation (Vaswani et al., 2017). To segment multiple sclerosis lesions on MRI data, a recent study proposed to train a model using soft masks to account for the high uncertainty in lesion borders' delineation (Kats et al., 2019). The soft masks were generated from the binary masks using morphological dilations. For the loss function, the authors used the soft version of the Dice loss (Milletari et al., 2016). This study reported an improved performance (+1.8% of Dice on the ISBI 2015 dataset) when using soft vs. binary masks. Another study suggested proposed another ground truth softening method using over-segmentation and smoothing based on the distance to an annotated boundary, and also reported better performance over hard labels (+0.7% of Dice on the MRBrainS18 dataset) (Li et al., 2020). However, according to the authors, the performance improvements were conditioned by optimizing some hyper-parameters (e.g., number of super-pixels, beta), suggesting a potential limitation to generalize to new datasets and tasks. In the studies of Li et al. and Kats. et al., alteration of ground truth was based on arbitrary modifications of the input mask (mathematical morphology) and might not truly represent the underlying PVE. Moreover, even if the network is fed with soft ground truths, this rich information somewhat vanishes down the line in the training pipeline by the use of sharp activation functions (e.g., sigmoid) and classification-based loss functions (e.g., Dice loss) (Deng et al., 2018; Jia et al., 2019).

## 1.2. Study outline

In this work, we explore training models using soft segmentations, both as input and output. While manual soft ground-truth generation is costly and highly time-consuming, we obtain soft inputs "for free" from binary ground truth data by skipping the binarization step that typically follows preprocessing and data augmentation. We focus on three key features: (i) training on soft (vs. hard) ground truth masks, (ii) the activation function used at the last layer (normalized ReLU vs. sigmoid), (iii) the use of a regression loss (vs. Dice loss) to favor soft predictions. We perform ablation studies for these three training features, whose combination is called SoftSeg, against the conventional training scheme on three open-source segmentation datasets: the spinal cord gray matter (SCGM) challenge (Prados et al., 2017), the multiple sclerosis (MS) brain lesion challenge (Commowick et al., 2018), and the multimodal brain tumor segmentation (BraTS) challenge 2019 (BraTS 2019). In the following sections, the differences between SoftSeg and the conventional training pipeline will be detailed, along with the evaluation framework we used to compare them. Second, the results of the comparison on the three datasets will be presented from different perspectives: (i) the training process, (ii) the qualitative aspect of the segmentation, and (iii) the quantitative performances. Finally, the key contributions of SoftSeg will be discussed as well as the perspectives it offers.

# 2. Material and methods

## 2.1. Proposed method

The comparison between a conventional training pipeline and our proposed approach, SoftSeg, is illustrated in Figure 1. The key differences involve the binarization of the input ground truth, the activation function, and the loss function. These differences are detailed in this section.

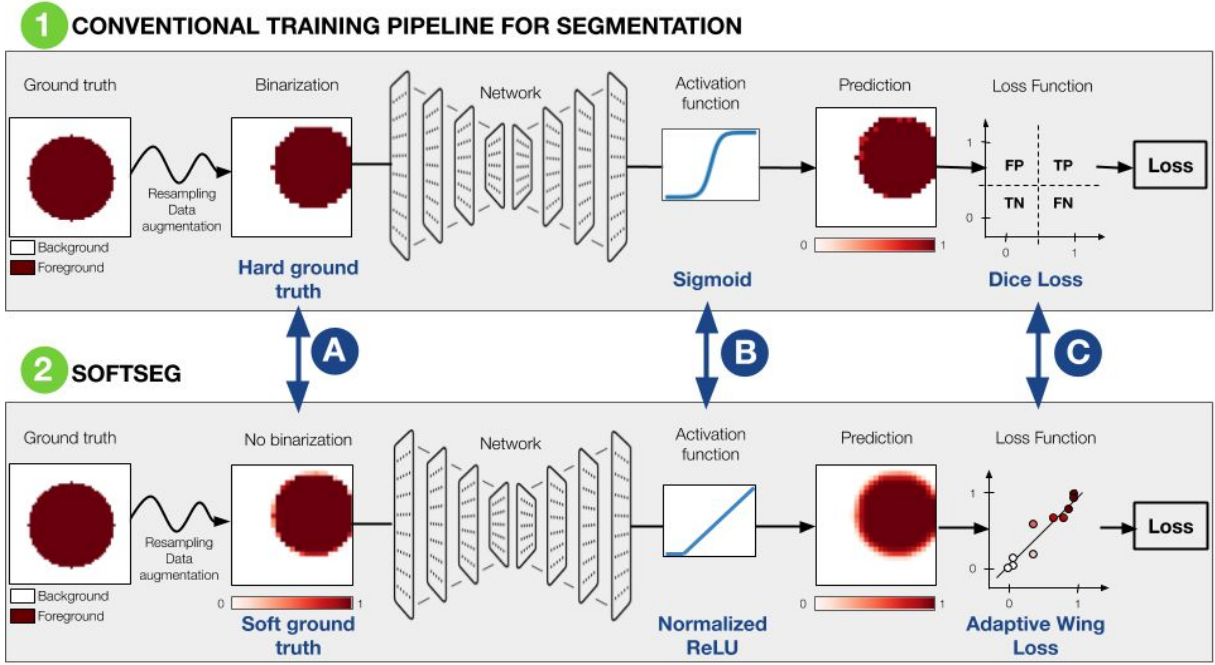

**Figure 1: Training pipelines for segmentation.** 1: Conventional training pipeline; 2: Our proposed approach (SoftSeg). The main differences are: (A) No binarization of the ground truth after the preprocessing and data augmentation operations; (B) A linear activation function is used instead of a sigmoid activation; (C) The loss function aims at solving a regression problem instead of a classification task.

### 2.1.1. Hard vs. Soft ground truth masks

Ground truth masks received by the network are conventionally binary, i.e., zeros and ones only, so-called "hard" ground truth. Although rarely specified, it is common to binarize the ground truth after applying preprocessing and data augmentation operations before feeding the network. Binarization is an approximation and a loss of information, especially for voxels at the border between two tissue types. To prevent such approximations, we propose to use soft (i.e., continuous values between 0 and 1) instead of hard masks, as illustrated in Figure 1A which are the result of the preprocessing and data augmentation without binarizing prior to the network. Soft masks used in this work notably aim at preserving partial volume information throughout the learning process, without applying complex label smoothing methods (Kats et al., 2019; Li et al., 2020) or resorting to costly soft ground-truths (e.g., from multiple experts).

### 2.1.2. Sigmoid vs. Linear activation function

The sigmoid activation function is popular in binary image segmentation models. Often used as the final activation, this non-linear activation is appropriate for classification since most values lie near 0 and 1, yielding a quasi-binary output. However, in the context of soft prediction, the sigmoid function undesirably narrows the range of soft values that potentially carry valuable PVE information. Although it can be partially addressed by increasing the temperature to make the active region larger, the use of other final activation functions (e.g., ReLU) has been recently explored, see for instance the comparison between CNN-Softmax and CNN-ReLU for classification tasks (Agarap, 2018). To avoid the polarizing effect in voxels observed when using the sigmoid, we propose to change the final activation of the segmentation model from the sigmoid function to a normalized rectified linear function (ReLU, see Figure 1B). A ReLU activation is applied to the model's output to set all negative values to 0 (V. Nair & Hinton, 2010). The result is then normalized by the maximum value to have a final output between 0 and 1, leading to a linear activation for the positive values and therefore highlighting the full range of prediction values from the model:

$$\text{NormReLU} \equiv \begin{cases} \frac{ReLU(X)}{max\{ReLU(X)\}}, & \text{if } \max\{\text{ReLU(X)}\} \neq 0 \\ 0, & \text{else} \end{cases}$$

where *X* represents the matrix output of the model before the final activation.

### 2.1.3. Classification vs. regression loss function

Segmentation is often considered a classification task where each voxel is assigned to one class. In that context, classification loss functions are commonly prioritized for segmentation tasks, such as the binary cross-entropy or the Dice loss functions. Although widely used with medical data (Gibson et al., 2018; Perone et al., 2018; Shen et al., 2018; Sudre et al., 2017), the Dice loss yields sharp segmentation edges (Deng et al., 2018; Jia et al., 2019), hindering predictions of non-binary values and can lead to a volumetric bias (Bertels et al., 2020). In contrast, the training approach we suggest is closer to a regression task in that the output prediction represents the input with high fidelity (e.g., an input voxel composed of 70% of the class of interest would produce an output prediction of 0.7). Consequently, we suggest using a regression loss function to train our network instead of a classification loss function (see Figure 1C). In this paper, we use the Adaptive Wing loss (Wang et al., 2019), which has shown fast convergence and efficient mitigation against class imbalance (Kaul et al., 2019).

## 2.2. Datasets

To compare our approach with the conventional pipeline, we selected three publicly-available datasets: the SCGM challenge (Prados et al., 2017), the MS brain lesion challenge (Commowick et al., 2018), and the multimodal BraTS challenge 2019 (BraTS 2019).

### 2.2.1. Spinal cord gray matter challenge

The SCGM dataset contains 80 MRI $T_2^*$-weighted 3D images of cervical spinal cord, evenly-acquired in four centers with different MR protocols and 3T scanners (Philips Achieva, Siemens Trio, Siemens Skyra). Demographics of the scanned subjects and acquisition parameters can be found in (Prados et al., 2017). The gray matter was manually segmented on each 3D image by four independent experts (inter-expert Dice score ranging from 89% to 93% when compared to majority voting). The binary ground truth used in our experiments was generated with voxel-wise majority voting across all four experts. The dataset totalizes 940 cross-sectional 2D slices, whose resolution varies across centers: from 0.25x0.25 $mm^2$ to 0.5x0.5 $mm^2$.

### 2.2.2. MS brain lesion challenge

The MS brain lesion dataset was presented during the MICCAI 2016 challenge. It includes MRI scans of 15 subjects with five contrasts: $T_1$-weighted, $T_1$-weighted Gadolinium-enhanced, T2-weighted, PD T2-weighted, and FLAIR. The data was evenly acquired from three different centers and scanners: Philips Ingenia (3T), Siemens Aera (1.5T), and Siemens Verio (3T). MS lesions were manually segmented by seven experts. A consensus segmentation obtained with the Logarithmic Opinion Pool Based STAPLE algorithm (Akhondi-Asl et al., 2014) is used as ground truth in our experiments. The Dice score fluctuates between 69% and 77% when comparing each expert segmentation with the consensus ground truth. Moreover, the resolution varies from one center to another: 1x0.5x0.5 $mm^3$, 1.25x1x1 $mm^3$, and 0.7x0.75x0.75 $mm^3$ (right-to-left, posterior-to-anterior, inferior-to-superior). The provided dataset was already preprocessed as follows: denoising with the non-local means algorithm (Coupe et al., 2008), rigid registration (O. Commowick et al., 2012) on the FLAIR contrast, brain extraction, and bias correction with N4 algorithm (Tustison et al., 2010).

### 2.2.3. BraTS challenge 2019

The BraTS challenge 2019 includes 335 subjects with high grade or low grade gliomas acquired from 19 different centers with varying acquisition protocols and 3T scanners (BraTS 2019). Four contrasts were provided: $T_1$-weighted, $T_1$-weighted Gadolinium-enhanced, T2-weighted, and FLAIR. The peritumoral edema, the Gadolinium-enhancing tumor, and the necrotic and non-enhancing tumor core were manually segmented by one to four expert neuro-radiologists according to a common protocol. Rigid registration to a common anatomical template, skull-stripping, and 1 mm isotropic resampling was performed on the provided dataset. 20 subjects with high grade gliomas were randomly chosen from the dataset to perform multiple trainings within a reasonable time, while allowing proper cross-validation between them. The 20 subjects selected are listed in the 'brats_subjects.txt' file (https://github.com/ivadomed/article-softseg). As our study focuses on the comparison between soft and hard segmentation, we did not perform multi-class training. Hence, a single label was retained for the experiments: the tumor core composed of the necrotic and enhancing tumor.

## 2.3. Training protocol

Different training protocols were selected for each dataset based on initial hyperparameter exploration (Table 1).

**Table 1: Training parameters for each dataset.** For all training parameters, please see configuration files: https://github.com/ivadomed/article-softseg/tree/main/config. Abbreviations: MS: multiple sclerosis; RPI: right-to-left, posterior-to-anterior, inferior-to-superior orientation; SCGM: spinal cord gray matter.

| | | SCGM dataset | Brain MS lesion dataset | BraTS dataset 2019 |
|---|---|---|---|---|
| **Preprocessing** | **Resample** | $0.25 \times 0.25 \times 2$ mm$^3$ (RPI) | 1 mm isotropic | 1 mm isotropic |
| | **Batch format** | 2D axial slices | | |
| | **Crop** | $128 \times 128$ pixels$^2$ | $160 \times 124$ pixels$^2$ | $210 \times 210$ pixels$^2$ |
| **Data Augmentation** | **Rotation** | $\pm$ 20 degrees | | |
| | **Translation** | $\pm$ 3% | | |
| | **Scale** | $\pm$ 10% | | |
| **Batch Size** | | 8 | 24 | 24 |
| **U-Net Depth** | | 3 | 4 | 4 |
| **Dropout Rate** | | 30% | | |
| **Learning Rate** | **Initial** | 0.001 | 0.00005 | 0.0001 |
| | **Scheduler** | Cosine Annealing | | |
| **Adaptive Wing Loss** | | $\epsilon$=1; $\alpha$=2.1; $\theta$=0.5; $\omega$=8 | | |
| **Early Stopping** | | Patience: 50 epochs ; $\epsilon$: 0.001 | | |
| **Maximum Number of Epochs** | | 200 | | |

### 2.3.1. Training / validation / testing split

For the SCGM challenge dataset, the four centers with their associated data were randomly split into groups of size two / one / one to compose the training, validation, and testing sets, respectively. We split the SCGM dataset according to the acquisition center to assess the approaches' ability to generalize to new acquisition parameters. For the MS brain lesion and BraTS segmentation tasks, we trained the networks on 60% of the patients, with 20% held out for validation and 20% for testing. Center-wise splitting was not possible for the MS brain or BraTS datasets as the origin of images was not directly available.

## 2.3.2. Preprocessing

All data were resampled to a common dataset-specific resolution (see Table 1), using spline interpolation (2nd order) for the images and linear interpolation for the ground truths. Cross-sectional slices were subsequently center-cropped to a common size specific to each dataset (see Table 1).

## 2.3.3. Data augmentation

For data augmentation, affine transformations were randomly applied to all training samples using linear interpolation (see Table 1 for details). Segmentation labels from the conventional approach (i.e., hard training) were binarized after applying data augmentation, while soft training candidates were untouched to preserve the softness of their augmented masks. We assessed the impact of binarized augmented masks (i.e., hard ground truth) compared to non-binarized augmented masks (i.e., soft ground truth), see section 2.4.1 for more details.

## 2.3.4. Intensity normalization

The intensities of each image were standardized by mean centering and standard deviation normalization. When several contrasts were available (MS brain, BraTS), this normalization was done on each contrast separately.

## 2.3.5. Iterations

All models were trained with a patience of 50 epochs and a maximum epoch count of 200. Batch sizes of 8, 24, and 24 were respectively used for the SCGM, brain MS, and BraTS datasets.

## 2.3.6. Optimization

The learning rate was modified throughout the training according to the cosine annealing scheduler with an initial value of 0.001 for the SCGM dataset, 0.0005 for the MS dataset, and 0.0001 for the BraTS dataset.

## 2.3.7. Network architecture

For all experiments, we used a U-Net architecture (Ronneberger et al., 2015) with a depth (i.e., number of downsampling layers) of 3 for the SCGM challenge and of 4 for the brain MS lesion challenge and BraTS data (see section 2.4.1 for details). The choice of depth was based on preliminary hyperparameters optimization. Batch normalization (Ioffe &

Szegedy, 2015), ReLU function, and dropout (Srivastava et al., 2014) followed each convolution layer. Convolution layers had standard 3×3 2D convolutions filters and a padding size of 1.

### 2.3.8. Activation function

Two different activation functions were tested on the model's output: sigmoid or normalized ReLU function (Figure 1-B). Throughout the experiments, we assessed the characteristics exhibited by the model's predictions when using either sigmoid or normalized ReLU.

### 2.3.9. Loss function

We compared the use of a regression loss function to a standard classification loss function for segmentation tasks (see Figure 1-C), using the Adaptive Wing loss (Wang et al., 2019) vs. the Dice loss (Milletari et al., 2016). The Adaptive Wing loss, initially introduced for heatmap regression for labeling facial key points, was chosen for its ability to propagate and predict soft values, but the proposed approach could work with other regression losses. For the Adaptive Wing loss, preliminary experiments led to the hyperparameters indicated in Table 1.

### 2.3.10. Implementation

Implementation and model training was done with ivadomed v2.2.1 (Gros et al., 2020). ivadomed is a Python-based open-source framework for deep learning applied to medical imaging (https://ivadomed.org/). To promote the reproducibility of our experiments, all configuration files can be found at https://github.com/ivadomed/article-softseg.

## 2.4. Evaluation

### 2.4.1. Evaluation protocol

To isolate the specific impact of each explored feature (hard/soft mask, activation function, loss), five candidates were compared (see Table 2). *Hard-Sig-Dice* represents the conventional deep learning candidate using binarization with a sigmoid activation function and Dice loss (Figure 1, panel 1). Our proposed hypothetically-best candidate is *Soft-ReLU-Wing* (Figure 1, panel 2, *SoftSeg*). *Hard-ReLU-Wing*, *Soft-ReLU-Dice*, and *Soft-Sig-Wing* each has only one feature changed from our proposed candidate.

Cross-validation was applied to each model candidate. For the SCGM datasets, each model was trained 40 times, with an even split on the test centers (10 trainings with center 1 as test set, 10 trainings with center 2 as test set, etc.). For the MS and BraTS datasets, each model was trained 10 and 15 times respectively, with a different dataset split for each model. For each of the evaluation metrics (see 2.4.2), a non-parametric 2-sided Wilcoxon signed-rank test compared the *Soft-ReLU-Wing* candidate with every other candidate. A p-value inferior or equal to 0.05 was considered significant.

**Table 2: Candidates description.** Each row represents a candidate (i.e. a training approach), whose features are detailed in the columns. Abbreviations: GT: ground truth.

| | Binary GT after Data Augmentation | Activation function | Loss function |
|---|---|---|---|
| *Hard-Sig-Dice* (Conventional) | Yes | Sigmoid | Dice |
| *Hard-ReLU-Wing* | Yes | NormReLU | Adaptive Wing |
| *Soft-Sig-Wing* | No | Sigmoid | Adaptive Wing |
| *Soft-ReLU-Dice* | No | NormReLU | Dice |
| *Soft-ReLU-Wing* (SoftSeg) | No | NormReLU | Adaptive Wing |

### 2.4.2. Evaluation metrics

Before computing the evaluation metrics, the network predictions were resampled to the native resolution (i.e., resolution of the native ground truth) and binarized. The threshold used to binarize the predictions was determined by searching for the optimal value (between 0 and 1 with an increment of 0.05) in terms of Dice score when using the trained model on the training and validation images. The metrics include: (i) Dice score, (ii) precision, (iii) recall, (iv) absolute volume difference (absolute volume difference between the ground truth and prediction, divided by the ground truth volume), (v) relative volume difference, and (vi) mean squared error (MSE). All metrics are expressed in percentages. For the MS lesion segmentation task, we also included lesion detection metrics which are clinically relevant: the lesion true positive rate (LTPR) and false detection rate (LFDR) as defined in (T. Nair et al., 2020). These detection metrics were not used for the other datasets (SCGM and BraTS), because in these cases there was always only one 3D target object per MRI volume.

## 3. Results

In the following sections, we compare how the features illustrated in Figure 1 influence the training process (section 3.1), the prediction values dynamic (section 3.2), and the overall model performance on the testing dataset (section 3.3).

## 3.1. Training process

Figure 2 shows the evolution of the training process across different model configurations. The conventional approach (*Hard-Sig-Dice*) yielded quasi-binary predictions from the very early stages of the training. Conversely, the other candidates produced predictions with low values on the gray matter surrounding at the early stages (see epoch #5 and 10), while at later stages the object is delineated with a soft segmentation (i.e., high prediction values within the object core and lower values on the edges). Among the three proposed training schemes (bottom rows), the candidate *Soft-ReLU-Dice* produced high prediction values (i.e., red voxels in Figure 2) earlier in the training process than the other two. Although the conventional candidate yielded high prediction values earlier, it did not necessarily trigger an "early-stopping" of the training earlier than the proposed candidates. The mean early stopping epochs were 123 and 128 for the conventional and the proposed approach, respectively. This means that training time was not importantly impacted when performing soft training. Unlike the output of the conventional candidate, the edges of the segmented object with soft training remained soft even at the final stages (see "Last epoch" in Figure 2). This was particularly the case for the *Soft-ReLU-Wing* candidate. Results of the *Soft-Sig-Wing* candidate are not depicted here because the model training did not converge during this experiment (see Table 3 for overall quantitative results).

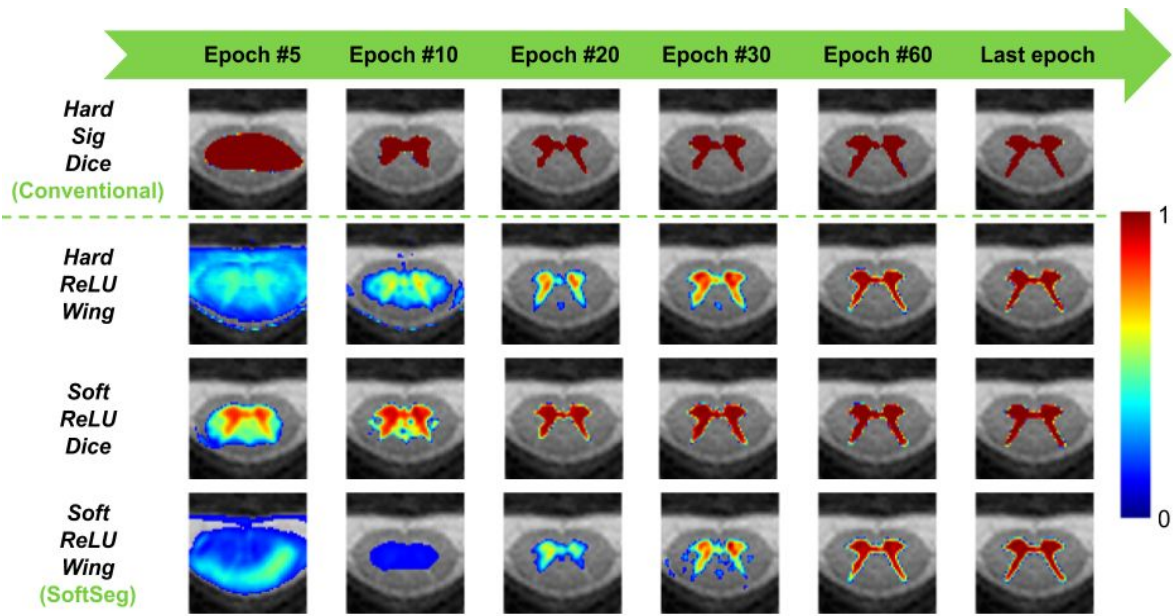

**Figure 2: Learning progression through epochs for different training schemes on the SCGM dataset.** Each row represents a training scheme, while each column shows the model prediction on a validation slice at a particular training epoch. The last epoch (right column) varied across approaches because of the early stopping feature. Predictions are overlaid on the anatomical data and range from 0 (transparent) to 1 (Red). *Soft-Sig-Wing* predictions are not shown here since the model training did not converge.

## 3.2. Output softness

To compare the performance of the different model configurations, we binarized the model predictions before computing the evaluation metrics. The binarization threshold was optimized by finding the value (between 0.05 and 0.95, with an incremental step of 0.05) that maximizes the Dice score when inferring on the training and validation dataset. Figure 3 shows the results of this optimization for each candidate (rows) and each iteration (purple dots). One notable observation is the large min-max Dice range across threshold values (up to 34% for *Soft-ReLU-Wing*), confirming the importance of this threshold optimization step. Conversely, the Dice range is more modest for the conventional candidate (9% for *Hard-Sig-Dice*), which is a direct consequence of the greater number of polarized values around 0 and 1. The loss function had the greatest impact on the min-max Dice range: it dropped from 34% to 13% when switching Adaptive Wing loss to Dice loss functions. This result highlights the importance of threshold fine-tuning when using a regression loss.

Figure 4 represents the voxels intensity distribution across the tested candidates and datasets. For the SCGM dataset (Figure 4A), all candidates yielded predictions with values concentrated around 1. *Soft-ReLU-Wing* intensity distribution is more spread out compared to other candidates and therefore its predictions could be considered being the least binarized. On the brain MS dataset (Figure 4B) and the BraTS brain tumor dataset (Figure 4C), two groups of candidates stand out: the "hard" group {*Hard-Sig-Dice*, *Soft-Sig-Wing*} and the "soft" group {*Soft-ReLU-Dice*, *Hard-ReLU-Dice*, *Soft-ReLU-Wing*}. In the "hard" group both candidates exhibit polarized predictions near 0 or 1. Conversely, the "soft" group values are more spread out in the ]0, 1] range (for the MS dataset) and ]0, 0.5] range (for the BraTS dataset). In the MS dataset, *Soft-ReLU-Wing* and *Soft-ReLU-Dice* are almost superimposed and yielded more non-zero values than *Hard-ReLU-Wing*. Across the three datasets, the "soft" group exhibits a higher number of non-zero predictions (higher area under curve).

Figure 5 illustrates the performance of each training scheme for the SCGM dataset in one representative subject per center. From this figure, one can appreciate the variability in terms of image resolution, white-to-gray matter contrast, and signal-to-noise ratio. Image heterogeneity had a notable impact on candidates' performance across test centers. On average, across all iterations, the candidates presented in Figure 5 obtained a Dice score of 86.2%, 81.2%, 88.6%, and 78.0% for centers 1, 2, 3, and 4, respectively. When compared with the conventional candidate, *Soft-ReLU-Wing* showed the highest Dice score for all test centers except for center #4 (77.4% for *Soft-ReLU-Wing* vs. 79.0% for *Hard-Sig-Dice*). Interestingly, in centers 1, 2, and 3 where images have the lowest resolution in the cross-sectional plane (0.3, 0.5, and 0.5 mm isotropic for centers 1, 2, and 3 respectively vs. 0.25 mm isotropic for center 4), the softer candidates segmented more truthfully the gray matter with an average improvement of 3.2% Dice score. This observation is in line with the hypothesis that soft training is well suited for mitigating PVE, i.e, the benefits are more considerable in images with lower spatial resolution.

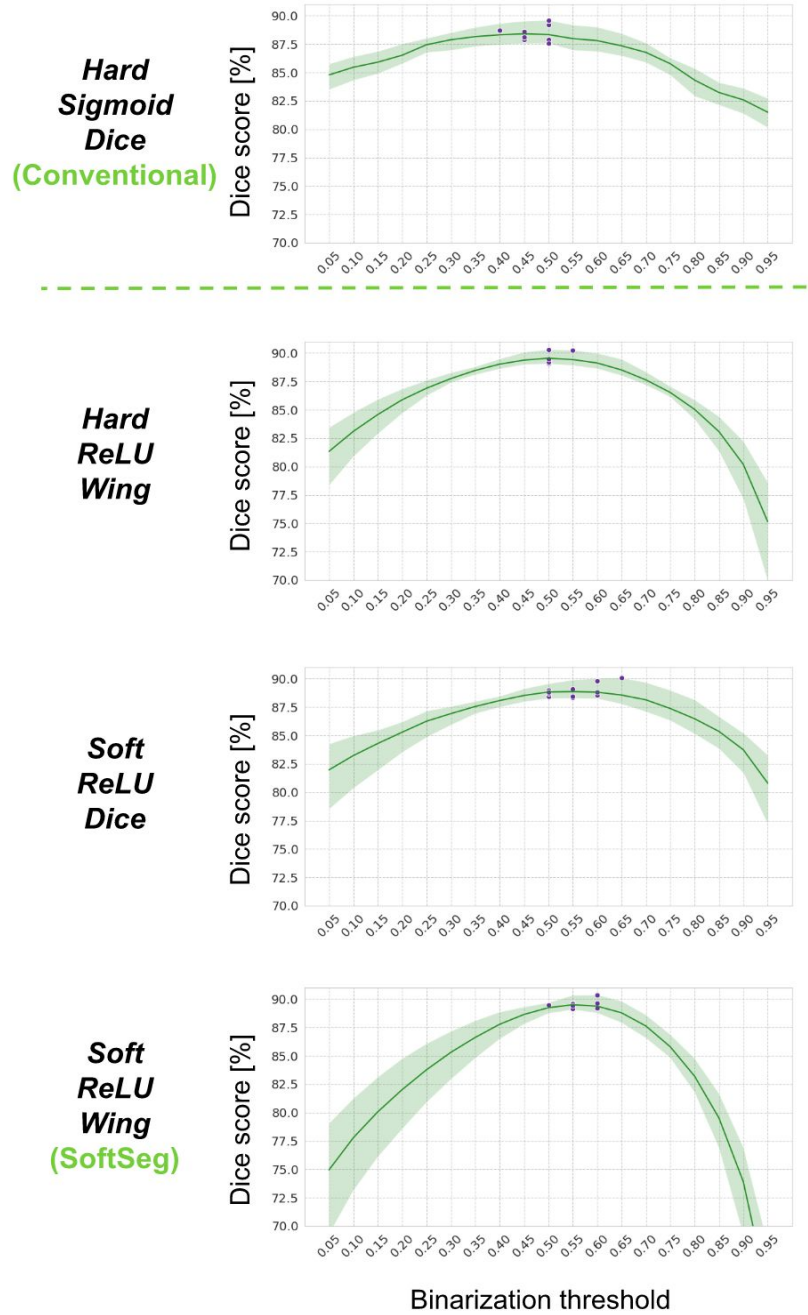

**Figure 3: Optimization of the binarization threshold for model prediction.** For each threshold value (between 0.05 and 0.95, with an incremental step of 0.05), the Dice score was computed on the trained model predictions for the training and validation SCGM data. The thick green line represents the average value while the green shaded area represents the min/max range of values. Purple dots represent the threshold that maximizes the Dice score, for each iteration. For the sake of comparison, the y-scale was kept the same across the four candidates. The lowest value for the *Soft-ReLU-Wing* (which is not shown due to cropping) is 56, and 70 for *Hard-ReLU-Wing*. *Soft-Sig-Wing* graph is not shown here since the model training did not converge.

**A. SCGM dataset**

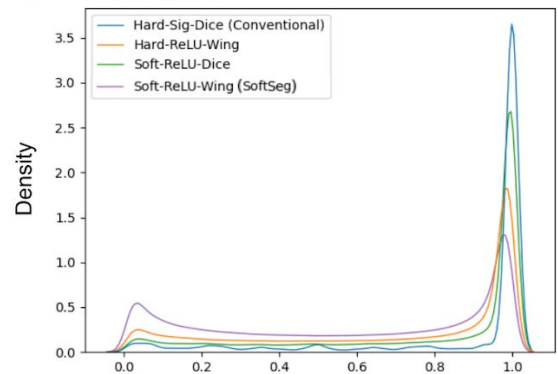

**B. Brain MS lesion dataset**

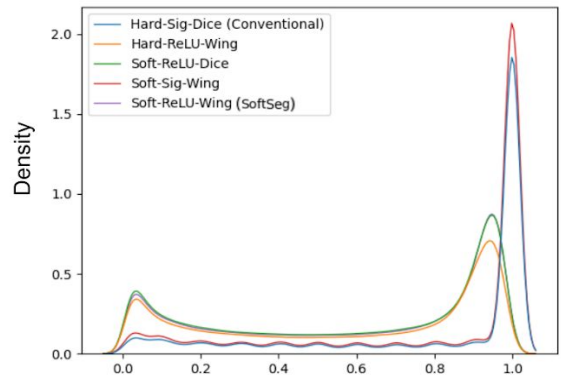

**C. Brain tumor dataset**

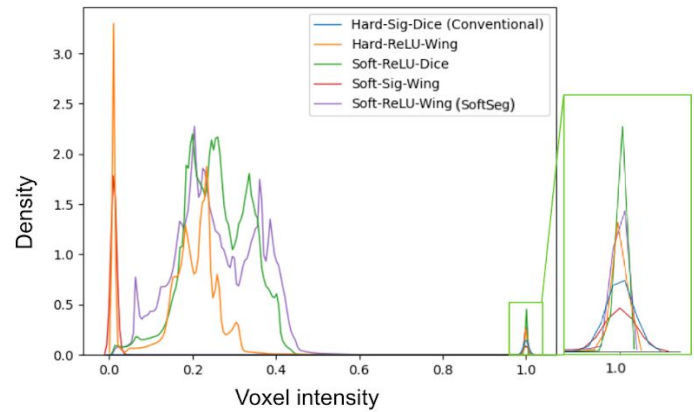

**Figure 4: Distribution of non-zero prediction voxels for each candidate on SCGM (A), MS brain lesions (B), and BraTS (C) datasets.** Distributions are computed using the kernel density estimation method and normalized so the area under the curve sums up to 1 for all curves. Training of the *Soft-Sig-Wing* model did not converge and is therefore not shown. The *Soft-ReLU-Dice* (green) and *Soft-ReLU-Wing* (purple) curves are almost perfectly superimposed on B. Because of the density estimation, the curves slightly extend outside of the prediction values (below 0 and above 1). Abbreviations: MS: multiple sclerosis ; SCGM: spinal cord gray matter.

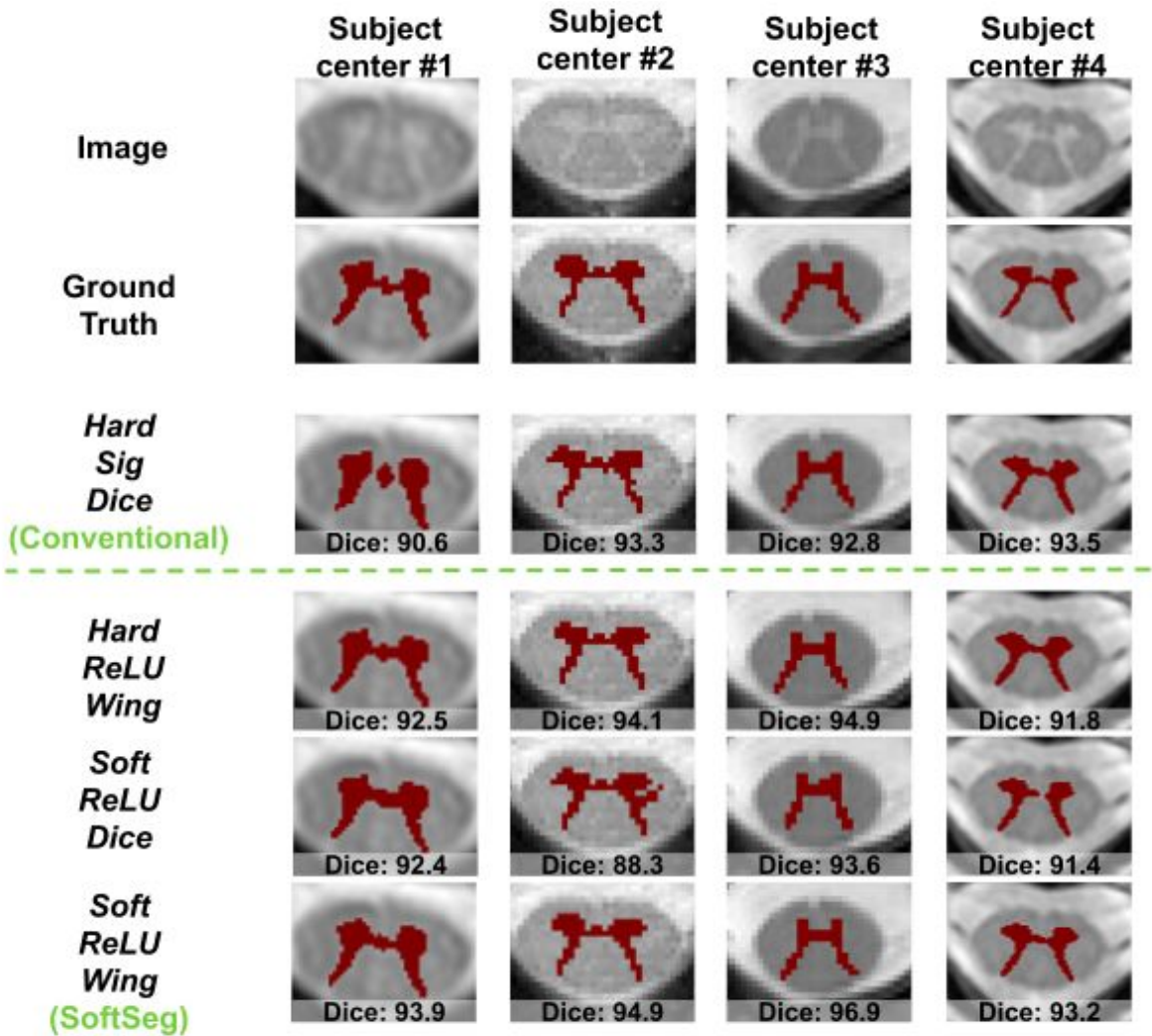

**Figure 5: Example of segmentation result for the SCGM dataset, across the four centers (columns) and the four candidates.** The first row shows the input 2D slice, the second row shows the manual ground truth. Rows 3-6 correspond to specific training schemes (see Table 2 for details). Predictions were binarized as described in section 2.4.2. Soft-Sig-Wing predictions are not shown here since the model training did not converge.

Figure 6 depicts MS lesion predictions across the five candidates. MS lesion predictions present two patterns of softness among approaches. *Hard-Sig-Dice* and *Soft-Sig-Wing* predict mostly values around 1 ("hard" group, as defined in in Figure 4 description), whereas *Soft-ReLU-Dice*, *Hard-ReLU-Dice*, and *Soft-ReLU-Wing* display a broader range of prediction values ("soft" group, as defined in Figure 4 description). The final activation distinguishes the two groups; the candidates displaying softer outputs had a normalized ReLU activation function, while the other candidates predicting more binarized values used a sigmoid as final activation. The candidates from the "hard" group, *Hard-Sig-Dice* and *Soft-Sig-Wing*, show overall less true positives (and consequently less false positives). Conversely, the softer candidates, *Soft-ReLU-Dice*, *Hard-ReLU-Dice*, and *Soft-ReLU-Wing*, are associated with a higher true lesions positive rate. On the close-ups from Figure 6 (left column), the candidates from the "hard" group show a single segmented lesion (two are missing), while all candidates from the "soft" group exhibit three distinct true positives.

Figure 7 illustrates segmentation results for the BraTS dataset. As observed in Figure 4 and 6, the same two groups with differing softness patterns can be isolated: *Hard-Sig-Dice* and *Soft-Sig-Wing* ("hard" group), and *Soft-ReLU-Dice*, *Hard-ReLU-Dice*, and *Soft-ReLU-Wing* ("soft" group). The "hard" group presents over-segmentation of the tumor core. Even on the false positive voxels, the raw prediction of the model yields a value of 1. Conversely, the soft group exhibits a ranging value of confidence around the borders of the tumor cores. The blue background on the *Soft-Sig-Wing* candidate is caused by most values being near 0 (not exactly 0). This candidate showed instability during training and did not reach convergence on every cross-validation. Like for the SCGM and MS lesion brain datasets, candidates from the "soft group" produce soft edges, and consistent shapes.

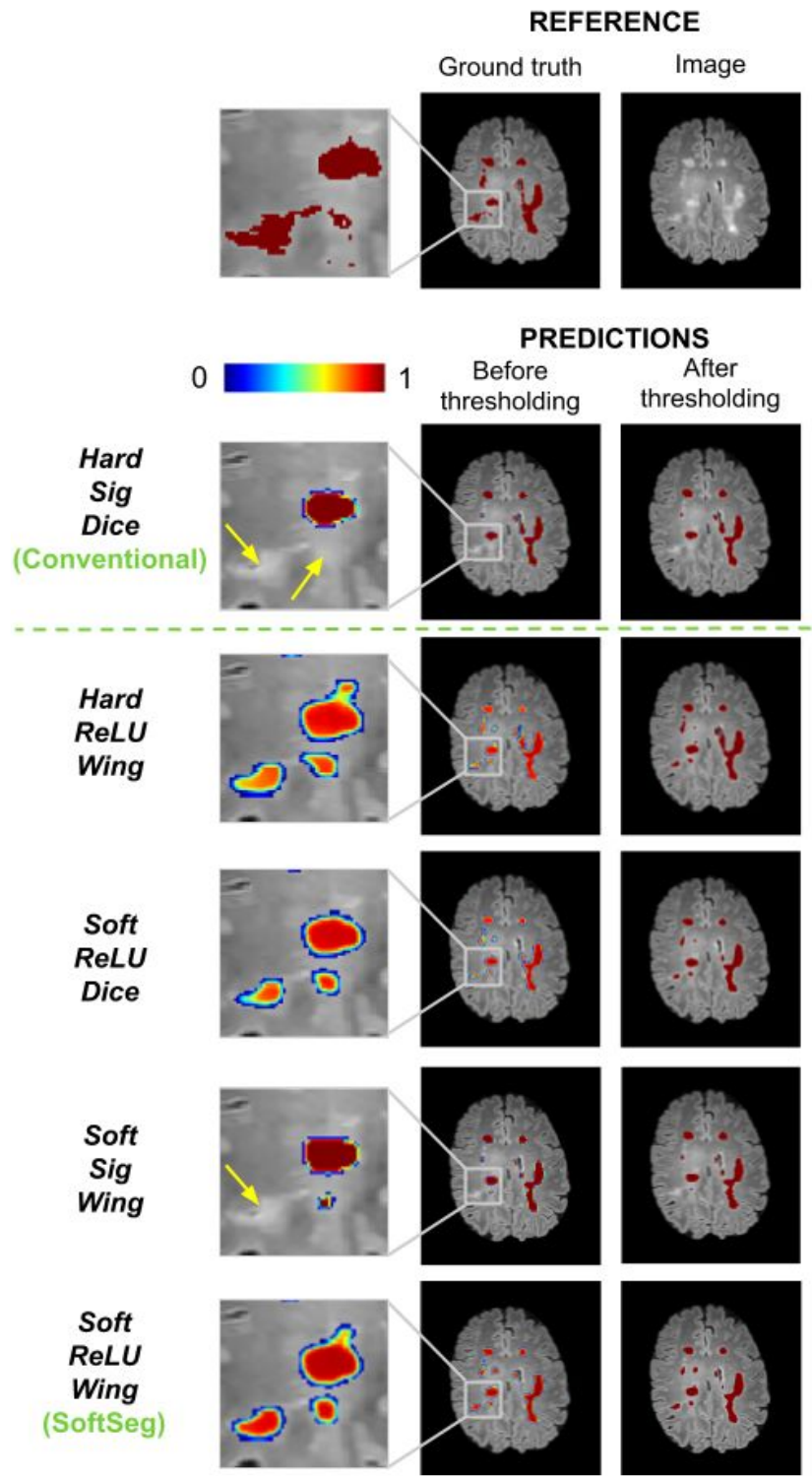

**Figure 6: Brain MS lesions segmentation for the five candidates.** The first row represents the input image and the consensus segmentation from the seven experts. For the remaining rows, the second column presents the raw predictions and the third column contains the binarized predictions. Predictions are overlaid on the anatomical data and range from 0 (transparent) to 1 (Red).

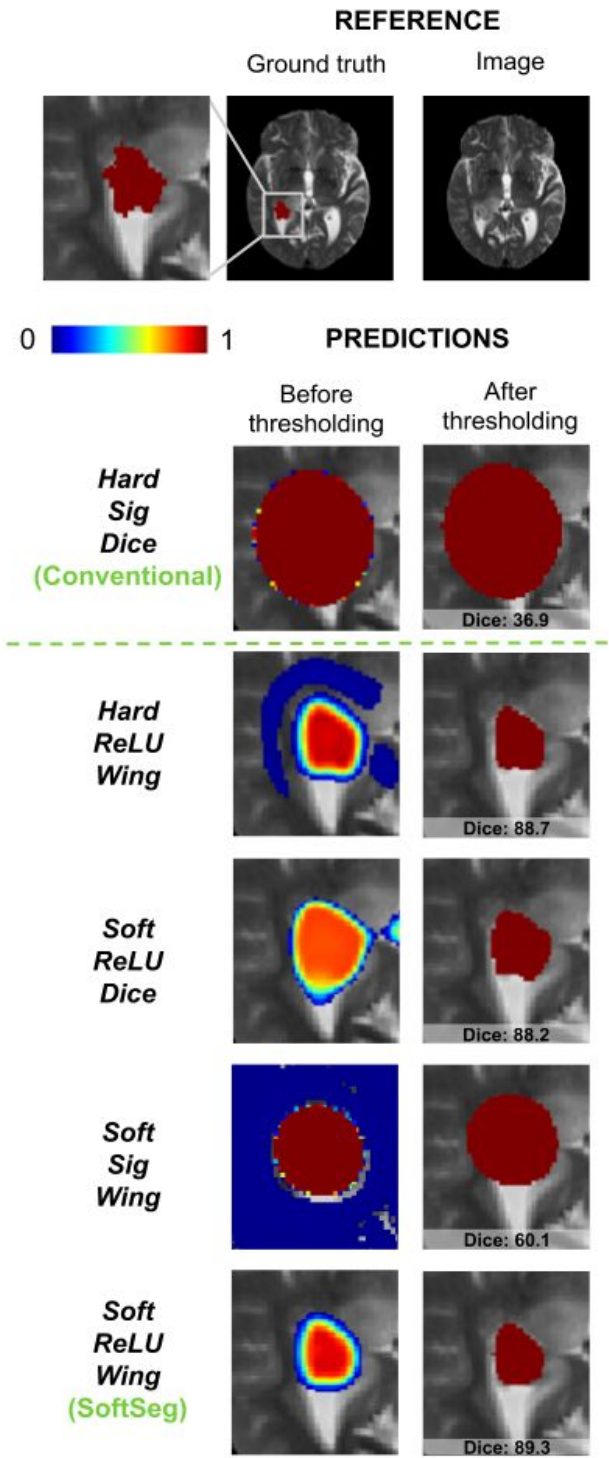

**Figure 7: Segmentation of brain tumor core for the five candidates.** The first row represents the input image and the ground truth with a close-up of the tumor segmentation. For the remaining rows, the left image represents the raw core tumor segmentation prediction from the model and the right the binarized prediction. Predictions are overlaid on the anatomical data and range from 0 (transparent) to 1 (Red).

## 3.3. Segmentation performance

### 3.3.1. SCGM

Table 3 summarizes the segmentation performance metrics for the five candidates on the SCGM dataset. *Soft-ReLU-Wing* yielded the highest Dice, precision, recall, absolute volume difference, and MSE scores compared to the conventional and other proposed approaches. When considering only the Dice score, there is a statistical difference between *Soft-ReLU-Wing* vs. *Hard-Sig-Dice* (p-value=0.0011), *Hard-ReLU-Wing* (p-value=0.0385), and *Soft-Sig-Wing* (p-value=1.10e-7). *Soft-Sig-Win*g did not converge with the GM dataset on all iterations which explains the low performances compared to the other candidates.

**Table 3: Gray matter segmentation performance metrics for the five candidates.** The error represents the standard deviation from 40 trainings (MEAN ± STD). The optimal score value is indicated under each metric name. Rows identify the five candidates (see Table 2 for candidates description). Columns represent the metrics.

** p-value < 0.05 for 2-sided Wilcoxon signed-rank test compared to the *Soft-ReLU-Wing* candidate. Abbreviations: MSE: mean squared error ; Opt: optimal.

| | Dice [%] *Opt. value: 100* | Precision [%] *Opt. value: 100* | Recall [%] *Opt. value: 100* | Absolute Volume Difference [%] *Opt. value: 0* | Relative Volume Difference [%] *Opt. value: 0* | MSE [%] *Opt. value: 0* |
|---|---|---|---|---|---|---|
| ***Hard-Sig-Dice*** (Conventional) | 82.3 ± 5.0 ** | 84.4 ± 9.2 ** | 83.3 ± 8.1 | 17.6 ± 11.7 | **-1.1 ± 20.3** | 0.290 ± 0.058 ** |
| ***Hard-ReLU-Wing*** | 83.7 ± 4.9 ** | 85.5 ± 11.3 | 84.5 ± 6.1 | 18.6 ± 13.8 ** | -2.0 ± 22.6 | 0.275 ± 0.082 |
| ***Soft-ReLU-Dice*** | 83.7 ± 5.2 | 85.7 ± 10.0 | 84.6 ± 7.6 | 17.6 ± 12.8 | -1.3 ± 21.2 | 0.269 ± 0.066 |
| ***Soft-Sig-Wing*** | 52.9 ± 36.7 ** | 73.1 ± 18.0 ** | 48.5 ± 39.1 ** | 53.0 ± 36.7 ** | 34.3 ± 54.3 ** | 0.611 ± 0.306 ** |
| ***Soft-ReLU-Wing*** (SoftSeg) | **84.3 ± 4.7** | **85.8 ± 10.8** | **84.9 ± 5.0** | **16.9 ± 12.1** | -1.6 ± 19.9 | **0.268 ± 0.083** |

### 3.3.2. Brain MS lesions

Table 4 presents the candidates performance metrics on the MS lesions dataset. As observed on the SCGM dataset, *Soft-ReLU-Wing,* had the highest Dice score, recall, and LTPR. *Soft-Sig-Wing* predicted less false positives compared to the other candidates illustrated by the highest precision score and the lowest LFPR. No statistical differences were observed between groups, probably due to large standard deviation between iterations on the MS dataset (testing set: n=3).

**Table 4: Brain MS lesion segmentation performance metrics for the five candidates.** The error represents the standard deviation from 10 trainings (MEAN ± STD). The optimal score value is indicated under each metric name. Rows identify the five candidates (see Table 2 for candidates description). Columns represent the metrics. Abbreviations: LFDR: lesion false detection rate ; LTPR: lesion true positive rate ; Opt: optimal.

| | Dice [%] Opt. value: 100 | Precision [%] Opt. value: 100 | Recall [%] Opt. value: 100 | LFDR [%] Opt. value: 0 | LTPR [%] Opt. value: 100 |
|---|---|---|---|---|---|
| ***Hard-Sig-Dice*** (Conventional) | 42.7 ± 14.5 | 58.3 ± 13.1 | 41.4 ± 17.1 | 61.9 ± 13.3 | 34.1 ± 16.7 |
| ***Hard-ReLU-Wing*** | 45.1 ± 13.0 | 55.5 ± 17.8 | 44.0 ± 15.5 | 65.6 ± 10.1 | 37.0 ± 14.4 |
| ***Soft-ReLU-Dice*** | 45.3 ± 14.1 | 56.6 ± 20.4 | 46.1 ±17.1 | 64.1 ± 12.0 | 36.5 ± 15.2 |
| ***Soft-Sig-Wing*** | 45.1 ± 12.2 | **59.8 ± 13.2** | 43.4 ± 13.2 | **57.7 ± 17.3** | 34.8 ± 15.6 |
| ***Soft-ReLU-Wing*** (SoftSeg) | **46.0 ± 12.2** | 55.2 ± 17.6 | **46.7 ± 13.8** | 63.0 ± 15.4 | **38.6 ± 14.8** |

### 3.3.3. Brain tumors

Table 5 reports the segmentation performance metrics of the candidates on the BraTS dataset. *Soft-ReLU-Wing* is associated with the highest Dice score, precision, relative volume difference, and MSE (Table 5). This candidate, when compared with the conventional candidate, reached statistical differences for precision (p-value=0.039), and MSE (p-value=0.024). The "soft" group, composed of *Soft-ReLU-Wing*, *Hard-ReLU-Wing*, and *Soft-ReLU-Dice*, presented similar Dice, precision, recall, MSE scores. *Soft-ReLU-Wing* yielded the highest recall score and *Hard-ReLU-Wing* the best absolute volume difference. As previously observed with the SCGM dataset, the candidate *Soft-Sig-Wing* did not converge on every iteration leading to lower segmentation scores. The conventional approach largely over-segmented tumor cores yielding an average relative volume difference of -29.9% and an average absolute volume difference of 67.1%, as illustrated in Figure 7.

**Table 5: Brain tumor segmentation performance metrics for the five candidates.** The error represents the standard deviation from 15 trainings (MEAN ± STD) on 20 randomly-selected subjects from the 2019 BraTS dataset. The optimal score value is indicated under each metric name. Rows identify the five candidates (see Table 2 for candidates description). Columns represent the metrics. ** p-value < 0.05 for 2-sided Wilcoxon signed-rank test compared to the *Soft-ReLU-Wing* candidate. Abbreviations: Opt: optimal.

|  | Dice [%] *Opt. value: 100* | Precision [%] *Opt. value: 100* | Recall [%] *Opt. value: 100* | Absolute Volume Difference [%] *Opt. value: 0* | Relative Volume Difference [%] *Opt. value: 0* | MSE [%] *Opt. value: 0* |
|---|---|---|---|---|---|---|
| *Hard-Sig-Dice* (Conventional) | 63.6 ± 28.7 | 66.1 ± 29.0 ** | 70.9 ± 30.8 | 67.1 ± 132.3 | -29.9 ± 145.6 | 40.1 ± 44.7 ** |
| *Hard-ReLU-Wing* | 57.2 ± 28.5 ** | 61.8 ± 32.0 ** | 70.0 ± 26.9 | 527.5 ± 164.3 ** | -490.2 ± 1654.8 | 92.2 ± 167.5 ** |
| *Soft-ReLU-Dice* | 69.8 ± 26.4 | **72.6 ± 28.8** | **73.2 ± 26.0** | 49.7 ± 83.1 | -20.7 ± 94.8 | 29.8 ± 35.0 |
| *Soft-Sig-Wing* | 55.7 ± 27.3 ** | 66.7 ± 30.3 | 60.1 ± 30.8 ** | 98.2 ± 193.5 ** | -45.9 ± 212.5 | 43.3 ± 43.8 ** |
| *Soft-ReLU-Wing* (SoftSeg) | **70.1 ± 23.2** | 71.9 ± 25.1 | 72.8 ± 25.0 | **38.6 ± 64.5** | **-8.1 ± 74.9** | **29.7 ± 38.0** |

# 4. Discussion

We introduced an alternative approach, SoftSeg, to train deep learning models for image segmentation. We demonstrate the application of SoftSeg in three different and publicly-available medical imaging datasets. The proposed training scheme is based on ground truth labels with continuous rather than binary values, in order to account for PVE and ill-defined object boundaries. These continuous labels are obtained for free as a side effect of non binarizing after data augmentation. To allow soft label propagation through the network training process, we modified the conventional training pipeline by using (i) soft ground truth masks, (ii) a normalized ReLU final activation layer, and (iii) a regression loss function (Adaptive Wing loss). Overall, the combination of these three features outperformed the conventional candidate on the three tested datasets (see Tables 3 to 5). Besides, this candidate yields soft predictions, especially at object boundaries or on small objects such as MS lesions. These soft predictions provide relevant insights on the model's confidence and allow meaningful automated post-processing. In particular, the proposed approach has an increased sensitivity (e.g., identify a higher number of lesions), which is desired by radiologists. The developed training pipeline is freely available as part of ivadomed (Gros et al., 2020).

## 4.1. Impact of the soft features for training

The three soft features differing from the conventional approach are a soft input, the final activation, and the loss function. These features had an overall positive impact on segmentation performance. Taken separately or altogether, they yielded the highest Dice score and best output softness for each of the three tested datasets. Removing one soft feature from the fully soft candidate (*Soft-ReLU-Wing*) slightly lowered the Dice score for the candidates that reached convergence. On the brain MS and BraTS datasets, the final activation had the greatest impact on the predictions' softness. Two different behaviors were clearly distinguishable when changing the final activation. The group associated with the normalized ReLU activation function ("soft" group) yielded softer predictions that can be assessed quantitatively (Figure 4) and qualitatively (Figure 6), when compared to the group with a sigmoid as final activation ("hard" group). In Table 4, the "soft" group can be associated with higher true positive detection rates (better recall and LTPR) and the "hard" group with less false positives (better precision and LFDR). This comparison cannot be made for the SCGM dataset, since the candidate with the conventional final activation did not converge. Similarly, the loss function and the use of hard vs. soft ground truths had overall a positive impact on the segmentation performance. Both features had an average Dice score drop of 0.8% across datasets when using their hard versions compared to the fully soft candidate, *Soft-ReLU-Wing*.

## 4.2. Non convergence of Soft-Sig-Wing

*Soft-Sig-Wing* performed poorly on SCGM and BraTS datasets (Tables 3 and 5). Some training runs from this candidate did not reach convergence while others did. Since *Soft-ReLU-Wing* always converged in our experiments, the instability of *Soft-Sig-Wing* may be attributed to the use of the sigmoid with a soft training approach. Since the sigmoid function tends to classify voxels (i.e., almost binary outputs), it may not be suitable to use in combination with a regression loss function which is not designed for polarized inputs. Consequently, the association of these two features could hinder training convergence.

## 4.3. Repeatability and statistical differences

Although rarely employed in deep learning model evaluation, we performed a cross-validation statistical analysis for each datasets. We used 40 folds on the SCGM dataset (10 per center), 10 folds on the brain MS dataset, and 15 folds on the BraTS dataset. For each dataset, the number of iterations for the cross-validation was determined by the typical training time while allowing us to run the different experiments in a reasonable time (~12 hours/training for the BraTS dataset on a single NVIDIA Tesla P100 GPU). Resorting to cross-validation to evaluate our approaches is particularly relevant for the brain MS and brain tumor datasets due to the small number of subjects. Also, the heterogeneity of lesion load and tumor core size led to high variations in performance across iterations (mean Dice standard deviation: 13.2% on MS lesions and 25.9% for brain tumors). Statistical difference was not reached for most metrics of MS and brain tumor dataset. The absence of statistical difference can be explained by the large standard deviations due to a wide performance range from one subject to another. The MS lesion and brain tumor datasets included 15 and 20 subjects leading to only 3 and 4 testing subjects respectively. Also due to the size of the dataset, only 10 (MS lesions) and 15 iterations (brain tumors) were performed on these datasets. Datasets with more patients leading to smaller standard deviations and more iterations would help in getting statistical differences.

## 4.4. Perspectives

### 4.4.1. Partial volume effect accountability

Multi-center studies consist of pooling datasets from different centers. This approach is hampered by the variability in acquisition parameters. In particular, variations in spatial resolution produce a varying amount of PVE, which in turn can hamper the precision and accuracy of morphometrics estimation (Chaves et al., 2020; Moccia et al., 2019). By preserving the PVE during data preprocessing (e.g., data resampling, template registration), a soft training approach has the potential to better quantify small changes in anatomical tissues. Future work could focus on evaluating the benefit of soft training against PVE variations in terms of biomarker quantification, such as MS lesion load quantification.

### 4.4.2. Encoding expert confidence during training

Manual segmentation of medical image segmentation is highly challenging and is prone to intra-expert variability. For instance, experts usually have a difficult time precisely delineating very small lesions (Carass et al., 2020). This challenge is partly due to them being required to decide if a voxel pertains to a lesion or not. The need for this binary decision has been driven by traditional training approaches, which require a binary ground truth as input for the model. With the SoftSeg method proposed here, expert raters will have the possibility to modulate their manual rating and assign values that reflect their level of confidence, e.g., 0.5, 1, and 2 would be respectively associated with a low, medium, and high confidence about the presence of a lesion. Although more time-consuming than binary manual segmentation, encoding expert confidence in neural networks via the generation of soft ground truths would likely have a positive impact on segmentation performance.

### 4.4.3. Preserving inter-expert variability

High inter-expert variability is a widespread challenge in medical image segmentation, resulting from factors such as image quality, expert training/experience (Carass et al., 2020; Shwartzman et al., 2019; Zhang et al., 2020). Some datasets provide the segmentation from multiple experts to account for this variability. However, these manual segmentations are usually merged into a single binary mask using label fusion methods, e.g., majority voting, STAPLE (Warfield et al., 2004) train deep learning models (Warfield et al., 2002). Recent studies highlighted the negative effect of label fusion methods to obtain reliable estimates of segmentation uncertainty as inter-expert variability is ignored when models are trained on the resulting binary masks (Jensen et al., 2019; Jungo et al., 2018). The SoftSeg method introduced here could elegantly account for the inter-expert variability and calibrate the model confidence by inputting soft ground truths that incorporate information about experts' disagreement. Future works could compare the benefit of using soft labels (obtained by averaging expert annotations) instead of hard labels (obtained from label fusion methods), both in terms of segmentation performance and uncertainty estimation.

### 4.4.4. Combining soft segmentation with uncertainty estimation

Estimation of deep learning model uncertainty is an active field of research (Camarasa et al., 2020; Loquercio et al., 2020; Mehta et al., 2020) in medical image segmentation, following the seminal works from (Gal & Ghahramani, 2016). Whether they are based on output probability calibration (Camarasa et al., 2020; Mehrtash et al., 2020), ensemble methods (Camarasa et al., 2020) or Bayesian models (T. Nair et al., 2020), all these approaches provide a representation of how trustful a prediction is. A common denominator of the recent investigations on uncertainty applied to segmentation tasks, is that they have relied on the conventional "hard" training, which produces highly polarized predictions, and as such might not be the most adequate for representing the rich spectrum of uncertainty values on the prediction. The conventional segmentation pipeline tends to yield overconfident predictions, even on misclassified voxels, leading to poor interpretation of the model's output (Mehrtash et al., 2020), which is well illustrated in Figure 7. A more comprehensive interpretation of deep learning model outputs would be achievable by estimating uncertainty on soft segmentation instead. Soft segmentation could also alleviate issues encountered with some uncertainty metrics which are sensitive to binary outputs (Camarasa et al., 2020).

Future investigations could evaluate the benefits of soft training used in combination with uncertainty estimation.

# 5. Conclusion

We presented SoftSeg, an approach to train deep learning models for image segmentation, which pursues the objective of soft predictions instead of the commonly-used hard outputs. In particular, SoftSeg can address the problem of partial volume effect by encoding (and propagating throughout the training chain) partial volume information in the ground truth masks. SoftSeg leads to informative and relevant soft outputs well calibrated while demonstrating an increase of performance on three open-source medical imaging segmentation tasks. Although used here with a simple 2D U-net as a proof-of-concept, SoftSeg can easily be integrated within already-existing deep learning architectures. Besides, SoftSeg could be leveraged to exploit a lossless combination of ground truth from multiple expert raters or to incorporate uncertainty estimation into an end-to-end soft framework.

# 6. Acknowledgements

The authors thank Joseph Paul Cohen, Olivier Vincent, Lucas Rouhier, Marie-Hélène Bourget, and Leander Van Eekelen for reviewing this manuscript. Funded by the Canada Research Chair in Quantitative Magnetic Resonance Imaging [950-230815], the Canadian Institute of Health Research [CIHR FDN-143263], the Canada Foundation for Innovation [32454, 34824], the Fonds de Recherche du Québec - Santé [28826], the Natural Sciences and Engineering Research Council of Canada [RGPIN-2019-07244], the Canada First Research Excellence Fund (IVADO and TransMedTech), the Courtois NeuroMod project and the Quebec BioImaging Network [5886, 35450]. C.G has a fellowship from IVADO [EX-2018-4], A.L. has a fellowship from NSERC and FRQNT. The authors thank the NVIDIA Corporation for the donation of a Titan X GPU.